\documentclass[showpacs,preprint,amssymb]{revtex4}
\usepackage{graphicx}
\usepackage{dcolumn}
\usepackage{bm}
\def\bi{\begin{itemize}}
\def\ei{\end{itemize}}
\def\bea{\begin{eqnarray}} 
\def\eea{\end{eqnarray}}
\def\be{\begin{equation}}
\def\ee{\end{equation}}
\def\line{\hbox to \hsize}    
\def\frac #1#2{{#1\over #2}}

\def\psid{\psi^{\dagger}}

\def \bz{{\overline z}}

\def\sgn{{\rm sgn\,}}

\def\ket #1{{\vert #1\rangle}}
\def\bra #1{{\langle #1\vert}}

\def\brak #1#2{{\langle#1, #2\rangle}}

\def\1{\mbox{\bf I}}

\def\cosec{{\rm cosec\,}}

\newenvironment{Quote}
{\begin{list}{}{%
\setlength{\leftmargin}{10 pt}
\setlength{\rightmargin}{\leftmargin}}
\item[]}
{\end{list}}

     {\refstepcounter{exercisenumber} \begin{Quote}\small{\sl
     Exercise~\thechapter.\theexercisenumber\/}:}
{\end{Quote}}

{\begin{Quote}\small{\sl Example\/}:}
{\end{Quote}}

\def\levelonelist{
        \begin{list}{\mybulA}%
                        {
        \setlength{\topsep}{0pt}
        \setlength{\parsep}{0pt}
        \setlength{\partopsep}{0pt}
        \setlength{\itemsep}{0pt}
                        }
                }
\def\leveltwolist{
        \begin{list}{\mybulB}%
                        {
        \setlength{\topsep}{0pt}
        \setlength{\parsep}{0pt}
        \setlength{\partopsep}{0pt}
        \setlength{\itemsep}{0pt}
                        }
                }

\def\el{\end{list}}

\usepackage{amsfonts}
\usepackage{amssymb}

\begin{document}

\title{An analogue of Hawking radiation in the quantum Hall effect}

\author{ MICHAEL STONE}

\affiliation{University of Illinois, Department of Physics\\ 1110 W. Green St.\\
Urbana, IL 61801 USA\\E-mail: m-stone5@illinois.edu}   

\begin{abstract}  
We use the identification  of  the edge mode of the filling fraction $\nu=1$ quantum Hall phase with   a 1+1 dimensional chiral Dirac fermion to construct an analogue  model for a chiral fermion in a space-time geometry possessing   an event horizon. By solving the model  in the lowest Landau level, we show that the event horizon emits particles and holes with a  thermal spectrum.  Each emitted quasiparticle  is correlated with an opposite-energy partner on the other side of the  event horizon. Once   we trace out these ``unobservable'' partners, we are left with a thermal density matrix. 

\end{abstract}


\maketitle

\section{Introduction}

 There are several apparently different  explanations for the origin of  black hole  radiation.  In his  original account  \cite{hawking74}  Hawking   kept track of what one means by a ``particle''  as  a wavefunction propagates in the background geometry.  A field theory derivation  using the trace anomaly in the energy momentum tensor was    given   given by Christensen and Fulling \cite{christensen-fulling}, and  more recently  Robinson and Wilczek \cite{robinson-wilczek} and others \cite{iso,banerjee-kulkarni}   have applied   the two-dimensional  gravitational anomaly in the region near the horizon.  Yet another route obtains the Hawking radiation from  quantum tunnelling across the horizon \cite{parikh-wilczek}.  (For a review of the tunnelling approach  see \cite{vanzo}.)

 Given these alternative derivations, it is reasonable to ask just  what is required  for an event horizon to emit thermal radiation. Is gravity really necessary? This question has led to the study of analogues of black holes and event horizons in other areas of wave propagation. The first such analogue was the 
acoustic black hole proposed  by Unruh, who discovered  that  the wave equation for sound in a background fluid flow was  equivalent to the wave equation for  a scalar field in a curved space-time \cite{unruh81}.  The subject has now developed extensively, with gravity and Hawking radiation analogues being proposed and constructed in quantum-fluids,  optics, and solid-state devices. For review with an extensive  list of references see \cite{visser_review}.  

The present  paper proposes a conceptually simple, and possibly experimentally realizable, condensed matter model   of quantum mode propagation  in which an event horizon emits thermal radiation. The analogue space-time is flat, but consists of two causally disconnected halves. It is therefore a member
of the general class of condensed-matter event horizons discussed by Volovik in \cite{volovik99}. Our model exploits the intepretation of the edge-modes of a filling fraction $\nu=1$  quantum Hall system as a massless chiral Dirac fermion whose local ``speed of light''  is determined by the potential that confines the   Hall fluid, and is therefore subject to external control. 

In the next section we describe the model in the language of first-quantized tunnelling. In the third section we adopt  a second-quantized formalism so as to obtain a  Bogoliubov transformation between two natural bases for the system. This allows us to display the physical  ``vacuum''  as a coherent superposition of particle-hole pairs that are entangled across the horizon.  Just as  the Minkowski pure-state vacuum is a thermal mixed state  when seen by a Rindler co-ordinate observer \cite{unruh76,unruh-review}, our pure-state vacuum appears thermal when we trace out the ``unobservable'' over-the-horizon member of each pair.

\section{Lowest Landau level}

We  model our black hole as  a  two-dimensional electron gas (2DEG) in the $\nu=1$ quantum Hall phase.   
We arrange for   the boundary of the 2DEG to lie  along the $y$ axis, with the region $x<0$ occupied by the gas, and the region $x>0$ empty. 
Now assume that we have engineered the ``confining'' potential $V$ to be of the   form 
\be
V(x,y)= \lambda xy,
\ee
and have chosen the  the direction of the perpendicular magnetic field $B$ so that  the  classical guiding-centre  drift velocity  is 
\bea
{\bf v}_{\rm drift}&=& \frac{1}{eB}\left(-\frac{\partial V}{\partial y},\frac{\partial V}{\partial x}\right)\nonumber\\
&=& \frac{ \lambda}{ eB} (-x,y).
\eea
The electrons move along   equipotentials $V(x,y)= E$, which, for this potential, are  rectangular hyperbol{\ae}\  that have the $x$ and $y$ axes as asymptotes.
In particular, the electrons at   the edge of our  2DEG move  vertically along the $y$ axis at velocity 
\be
v_{\rm edge}= \frac{1}{eB}\frac{\partial V}{\partial y}=\frac{\lambda}{eB} y.
\ee
This  velocity is our  analogue of the local speed of light.
The  edge modes in   the regions $y>0$ and $y<0$ move in opposite directions, and so these two regions are    causally disconnected. They are separated by an event horizon at $y=0$.  

The price we pay  for the event horizon  is that the electrons in the occupied region with $y<0$ (the interior of the black hole) are  in a state of population inversion. In the absence of the magnetic field the electrons  would rapidly fall into  one of the the lower energy quadrants. Because of the strong  field, however, and in the absence of inelastic  or tunnelling processes, they are constrained to stay on their hyperbolic classical orbits.   
The inherent  instability of the ``vacuum'' in the  black hole interior corresponds to the observation of  Parikh and Wilczek  \cite{parikh-wilczek} that a black hole must be thought of as highly excited quantum state.

\begin{figure}
\includegraphics[width = 5.0 in]{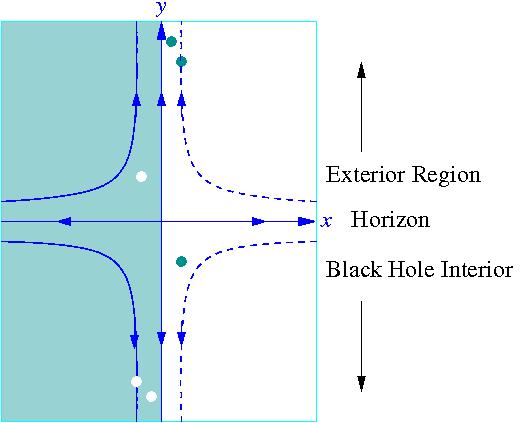}
\caption{\sl The 2DEG black-hole analogue. The shaded region is the 2DEG. The  lines indicate the  semiclassical electron orbits (dashed when mostly unoccupied). The low energy  excitations near the boundary at $x=0$ constitute the quantum system in which we will find Hawking radiation. This  radiation is illustrated by three  correlated pairs of electrons and holes moving in opposite directions inside and outside the black hole. }
\label{FIG:potential}
\end{figure}

Except for the case $E=0$, each  of the classical equipotentials $\lambda xy= E$ consists of two disconnected branches and intitally all the particles lie on only one of these branches.  The branches for small $E$ approach each other near  the origin. There is therefore   a non-zero amplitude for a particle to tunnel from one branch to the other of the same energy. This tunnelling  leads to electrons and holes being emitted from the event horizon. 

To calculate the tunneling amplitude, we will   assume that the magnetic field is large enough that we can ignore all Landau levels except the lowest. The lowest Landau level  (LLL) approximation is  very natural   as   it is this situation that the excitations near  the edge of a  quantum Hall droplet can be identified with those of a 1+1 dimensional chiral fermion with Hamiltonian
\be
\hat H= \int_{-\infty}^\infty v_{\rm edge}(y)\hat \psid(-i\partial_y) \hat \psi\,dy.
\ee
In this picture the  2DEG itself is   the  filled Dirac sea.

We chose the symmetric gauge in which  the  LLL wave-functions are of the form
\be
\psi(x,y) = \exp\left\{-\frac 14 eB|z|^2\right\} \psi(z),
\label{EQ:LLLfunction}
\ee
where $z=x+iy$.   All  quantum information resides in the holomorphic factor $\psi(z)$,  and  we will  refer to this factor  as  the LLL ``wave-function.''  We therefore  regard the LLL Hilbert space as a Bargmann-Fock space of finite-norm holomorphic functions with inner product 
\be
\brak{\varphi}{\chi}= \int d^2 z \, e^{-eB|z|^2/2}\, \overline{\varphi(z)}\chi(z), \quad d^2z \equiv  \frac 1 {2 i} d\bz\wedge dz= dx\wedge dy. 
\ee
Bear in mind however that the LLL wavefunction should  be multiplied by  $\exp\left\{-\frac 14 eB|z|^2\right\}$ before plotting probability densities  or computing currents.

The  action of $z$ on the LLL wavefunction is by simple multiplication,  but multiplication by   $\bz$ takes us out of the space of holomorphic functions.   The LLL  operator corresponding to $\bz$  becomes instead    
 $z^\dagger$, where  the adjoint is  taken with respect to the Bargmann-Fock inner product. This identification makes 
 \be
\bz    \mapsto z^\dagger =\frac{2}{eB}\frac{\partial}{\partial z}.
\ee

For our potential 
\be
 \lambda xy = \frac{\lambda}{4i}(z^2-\bz^2)
\ee
the first-quantized eigenvalue problem
\be
H\psi=\epsilon\psi
\ee
therefore  becomes
\be
\left(\frac{1}{e^2B^2}\frac{d^2}{dz^2}-\frac {z^2}4\right)f(z)=- i \frac{\epsilon}{\lambda}f(z).
\ee
Only the potential appears in the this equation as the LLL wavefunctions are annihilated by the electron kinetic energy operator.
A   rescaling gives us a standard form of Weber's equation (See \cite{whittaker-watson} \textsection16.5, or \cite{abramowitz-stegun} chapter 19.):
\be
\left(\frac{d^2}{d \zeta^2}-\frac {\zeta ^2}{4}\right)f(\zeta )=a f(\zeta),
\label{EQ:originalweber}
\ee
with
\be
a= -{i\epsilon}\left(\frac{eB}{\lambda}\right)f(\zeta),\quad \zeta= \sqrt{eB}z=\frac{z}{\ell_{\rm mag}}.
\ee
For simplicity we  now set $\lambda=eB=1$.  We can always restore the general parameters by  scaling the units of length and energy.  

If $\varphi(\epsilon, z)$ is a solution of
\be
\left(\frac{d^2}{d \zeta^2}-\frac {\zeta ^2}{4}\right)\varphi(\epsilon, \zeta )=-{i\epsilon}\varphi(\epsilon, \zeta)
\label{EQ:ourweber}
\ee
then so are $\varphi(\epsilon, -z)$, $\varphi(-\epsilon, iz)$ and $\varphi(-\epsilon,-iz)$. At most two of these solutions can be linearly independent.

\begin{figure}
\includegraphics[width=2.5in]{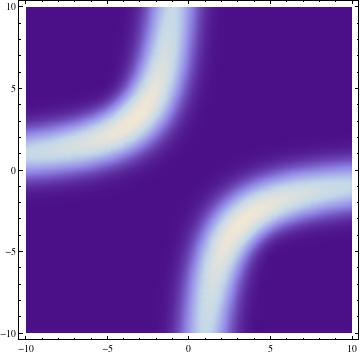}
\includegraphics[width=2.5in]{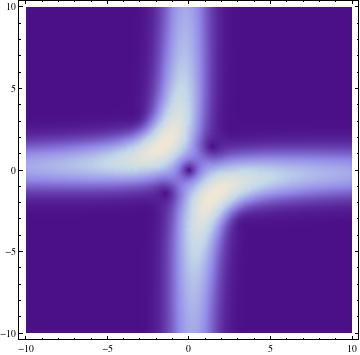}
\caption{\sl Left figure:  A density plot of the the absolute value of the even function $\exp\{-|z|^2/4\}y_1(x,y)$ for the case $\epsilon=-10$. Right figure: A density plot of   the absolute value of the odd function $\exp\{-|z|^2/4\}y_2(x,y)$ for $\epsilon=-2$.}
\label{FIG:y1}
\end{figure}

A fundamental  pair of independent solutions is
\bea
y_1(\epsilon, z) &=&e^{-z^2/4}{}_1F_1\left(\frac 14- i\frac{\epsilon}{2}, \frac 12, \frac{z^2}{2}\right),\nonumber\\
y_2(\epsilon,  z) &=&ze^{-z^2/4}{}_1F_1\left(\frac 34-i \frac{\epsilon}{2}, \frac 32, \frac{z^2}{2}\right).
\eea
Here ${}_1F_1(a,b,z)$ is the confluent hypergeometric function.
These functions are even and odd, respectively, under $z\leftrightarrow -z$.  After multiplication by $\exp\{-|z|^2/4\}$ the resulting wavefunctions are  localized on the semiclassical orbits, which form the two disconnected branches of the rectangular hyperbola $xy=\epsilon$ (see figure \ref{FIG:y1}).
These solutions to the LLL potential have been studied in connection with Riemann hypothesis \cite{townsend}.

More useful to us is the  solution of (\ref{EQ:ourweber}) given by the  parabolic cylinder function 
\bea
U_{-i\epsilon}(z)&\equiv& D_{i\epsilon-1/2}(z)\nonumber\\
&=& 2^{-(i\epsilon/2+1/4)}\frac{1}{\sqrt{\pi}}\left(\cos\left[\pi \left(\frac 14 -\frac{i\epsilon}{2}\right)\right]\Gamma\left(\frac{1}{4}+\frac{i\epsilon}{4}\right)y_1(\epsilon,z)\right.\nonumber\\
&& \qquad \left. -\sqrt{2} \sin\left[\pi \left(\frac{1}{4}-\frac{i\epsilon}{2}\right)\right]\Gamma\left(\frac{3}{4}+\frac{i\epsilon}{2}\right)y_2(\epsilon,z)\right)\nonumber\\
&=&\frac{e^{-z^2/4}}{ \Gamma(\frac 12-i\epsilon)}\int_0^\infty t^{-i\epsilon-\frac 12}e^{-\frac 12 t^2-zt}\,dt.
\label{EQ:cylinderdef}
\eea
Here  $D_n(z)$ is Whittaker and Watson's notation for their parabolic cylinder function \cite{whittaker-watson} , and $U_n(z)$ is the now more common notation used by Abramowitz and Stegun \cite{abramowitz-stegun}.  The essential properties of $U_n(z)$ are  that it is an entire function, and that it decays rapidly as $x\to +\infty$ for any real or complex $n$.  

\begin{figure}
\includegraphics[width=3.0in]{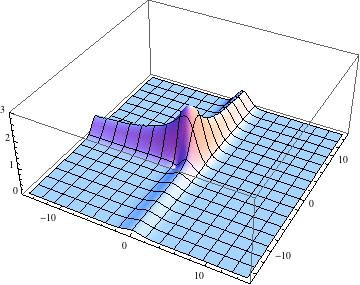}
\includegraphics[width=2.5in]{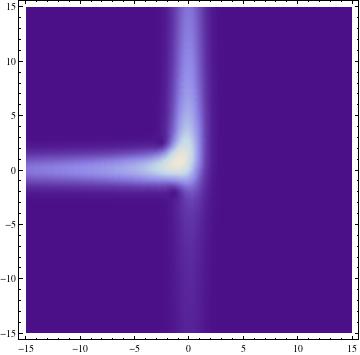}
\caption{\sl A countour and a density  plot of  $\exp\{-|z|^2/4\}U_{-i\epsilon}(z)$   for the case $\epsilon=-.5$.  Particles  enter from the left  and the beam divides between  down-going  and weaker tunnelled edge-mode wave  and the up-going and stronger direct edge--mode wave. }
\label{FIG:tunnel1}
\end{figure}

The solution $U_{-i\epsilon}(z)$ describes particles moving in from the left (the occupied region) in the lower left quadrant if $\epsilon>0$ and the upper left quadrant if $\epsilon<0$.  They mostly remain in  that quadrant, but there is some probability of tunnelling to the other branch of the hyperbola (see figure \ref{FIG:tunnel1}). If $\epsilon>0$ the  result is that a tunnelled positive energy particle  is emitted by the black hole,  leaving a negative energy hole ({\it i.e.\/}\ the absence of positive energy particle) inside the event horizon. If $\epsilon<0$ then a positive energy hole (the absence of a negative energy particle)  is emitted by the black hole leaving a negative energy particle inside the event horizon.

Along with the solution $U_{-i\epsilon}(z)$ we have the solutions $U_{-i\epsilon}(-z)$ and $U_{i\epsilon}(iz)$ and 
$U_{i\epsilon}(-iz)$. We will find use for all of these solutions, as they describe motion with different boundary conditions (see figures \ref{FIG:instates} and \ref{FIG:outstates}).

To discover the relative amplitudes of the direct and tunnelled waves we can use the asymptotic expansion 
\be
e^{-|z|^2/4}U_{-i\epsilon}(z)\sim e^{-|z|^2/4-z^2/4}z^{i\epsilon -1/2}\left[1+O\left(\frac 1{z^2}\right)\right], \quad |\arg(z)|<3/4.
\ee
Near the $y$ axis this reduces to  
\be
 \psi(x,y) \sim \hbox{(gauge phase)} e^{-x^2/2} \frac{1}{\sqrt{y}} \exp\{i\epsilon \ln|y|-\sgn(y) \epsilon \pi/2\}.
\ee
The ratio of tunneled to direct amplitude is therefore exactly $\exp\{-\pi \epsilon\}$.

We can conform this result by using the identity  
\be
U_{-i\epsilon} (z) = \frac{\Gamma\left(\frac 12 +i\epsilon\right)}{\sqrt{2\pi}} \left[e^{-\epsilon \pi/2}e^{-i\pi/4}U_{i\epsilon} (iz)+e^{\epsilon\pi/2} e^{+i\pi/4} U_{i\epsilon}(-iz)\right]
\ee
together with the fact that $U_{-i\epsilon}(z)$ tends rapidly to zero in the right half-plane.  Thus, if $R$ is positive
\bea
U_{-i\epsilon} (iR)&=&  \frac{\Gamma\left(\frac 12 +i\epsilon\right)}{\sqrt{2\pi}} \left[e^{-\epsilon \pi/2}e^{-i\pi/4}U_{i\epsilon}(-R)+e^{+\epsilon\pi/2} e^{+i\pi/4} U_{i\epsilon}(R)\right]\nonumber\\
&\sim& \frac{\Gamma\left(\frac 12 +i\epsilon\right)}{\sqrt{2\pi}}e^{-\epsilon \pi/2}e^{-i\pi/4}U_{i\epsilon}(-R)
\eea
and
\bea
U_{-i\epsilon} (-iR)&=&  \frac{\Gamma\left(\frac 12 +i\epsilon\right)}{\sqrt{2\pi}} \left[e^{\epsilon \pi/2}e^{-i\pi/4}U_{i\epsilon}(R)+e^{\epsilon\pi/2} e^{+i\pi/4} U_{i\epsilon}(-R)\right]\nonumber\\
&\sim& \frac{\Gamma\left(\frac 12 +i\epsilon\right)}{\sqrt{2\pi}}e^{\epsilon \pi/2}e^{+i\pi/4}U_{i\epsilon}(-R).
\eea
The direct and tunneling amplitudes  therefore have magnitude 
\bea
|d(\epsilon)|&=&  \left| \frac{\Gamma\left(\frac 12 +i\epsilon\right)}{\sqrt{2\pi}}\right|e^{\epsilon \pi/2}\nonumber\\
|t(\epsilon)|&=&  \left| \frac{\Gamma\left(\frac 12 +i\epsilon\right)}{\sqrt{2\pi}}\right|e^{-\epsilon \pi/2}
\eea
Note that $|d(\epsilon)|^2+|t(\epsilon)|^2=1$ because $\Gamma(z)\Gamma(1-z)=\pi \cosec (\pi z)$ gives us
\be
 \left|\Gamma\left(\frac 12 +i\epsilon\right)\right|^2=\frac{2\pi }{ e^{\pi \epsilon}+e^{-\pi \epsilon}}.
\ee

The occupation  probability of  an outgoing   particle or hole   state with energy $\epsilon$ is therefore
\be
P(\epsilon)= \frac{1}{1+\exp\{2\pi \epsilon\}}.
\ee
The chiral edge states emerging from the event horizon are  therefore thermal with
\be
T= \frac 1{2\pi},
\ee
or 
\be
k_BT=\frac{\hbar \lambda}{2\pi e B},
\ee
once we restore parameters and units. Comparison of this with the usual Hawking radiation formula 
\be
k_B T_{\rm Hawking}= \frac{\hbar \kappa}{2\pi}
\ee
indicates that the ``surface gravity'' $\kappa$ of our analogue black hole is the edge-velocity  acceleration 
\be
\kappa= \frac{\lambda}{eB}=\left. \frac{d v_{\rm edge}}{dy}\right|_{\rm horizon}.
\ee.

\section{Second quantization, mode expansions,  and a Bogoliubov tranformation}

The space of LLL functions  ({\ref{EQ:LLLfunction}) does not contain the delta function. Its place is taken by a reproducing kernel  
\be
 \{x_1,y_1|x_2,y_2\} \stackrel{\rm def}{=} \frac 1{2\pi} \exp\left\{-\frac 14 |z_1|^2-\frac 14 |z_2|^2/4 +\frac 12 \bz_1 z_2\right\}.
 \ee
If $\psi(x,y)$ is of the form ({\ref{EQ:LLLfunction})  then
\be
\int d^2 z_1  \psi (x_1,y_1)   \{x_1,y_1|x_2,y_2\}= \psi(x_2,y_2).
\ee
In particular $\{x_0,y_0|x_1,y_1\} $, considered  as a function of $(x_1,y_1)$, is  of this form, so  the kernel reproduces itself:
\be
\int d^2 z_1  \{x_0,y_0|x_1,y_1\} \{x_1,y_1|x_2,y_2\}= \{x_0,y_0|x_2,y_2\}.
\ee
When  we expand out the second quantized LLL field operator  in terms of the  discrete set of normalized eigenmodes $z^n/\sqrt{2\pi 2^n n!}$ for the potential 
\be
V(x,y)=\frac 12  (x^2+y^2) = \frac 12 z \bz \mapsto z\frac{d}{dz}
\ee 
we find 
\be
\widehat \psi(x_1,y_1) = \sum_{n=0}^\infty \widehat a_n \frac{1}{\sqrt{2\pi 2^n n!}} z^n e^{-|z|^2/4}.
\ee
Here the  operators $\widehat a_n$ obey  
\be
\{ \widehat a_n,\widehat a^\dagger_m\}=\delta_{nm},
\ee
and  the usual canonical anticommutation relation $\{\widehat \psid(x),\widehat \psi(x')\}= \delta(x-x')$  for the field is  replaced by 
\be
\{\widehat \psi^\dagger(x_1,y_1 ),\widehat\psi(x_2,y_2)\}= \{x_1,y_1|x_2,y_2\}.
\ee
If we retain only the holomorphic factors,  then we have
\be
\{\widehat \psi^\dagger(z_1),\widehat \psi(z_2)\}= \frac 1 {2\pi} \exp\left\{\frac 12 \bz_1 z_2\right\}
\label{EQ:holomorphic_anti}.
\ee 

We can also expand in a continuous set of  continuous set of eigenfunctions. For example,  we can make use of energy $E$   eigenfunctions for the potential  $V(x,y)=x$.
These are 
\be
\varphi_E(z)=\frac{1}{\pi^{1/4}}\exp\left\{Ez-\frac 14 z^2 -\frac{1}{2}E^2\right\}.
\ee
They have been normalized so that 
\be
\brak{\varphi_E}{\varphi_{E'}}= 2\pi\,\delta(E-E').
\ee 
The holomorphic  field operator is then
\be
\widehat \psi(z) = \int_{-\infty}^{\infty}\frac{dE}{2\pi} \,\widehat a_E \varphi_E(z)
\label{EQ:linear_expansion}
\ee
with 
\bea
\{\widehat a_E,\widehat a_{E'}^\dagger\}= 2\pi\, \delta(E-E').
\eea
We easily confirm  that (\ref{EQ:linear_expansion}) still satisfies (\ref{EQ:holomorphic_anti}).

\begin{figure}
\includegraphics[width=4.5in]{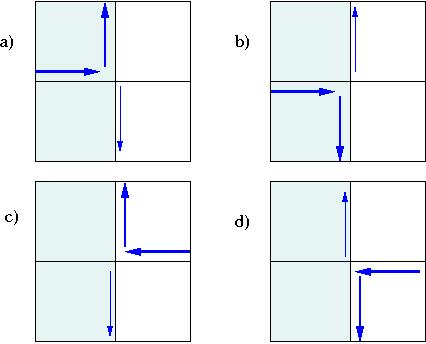}
\caption{\sl The ``in'' wavefunctions: a) $\varphi^{({\rm in},+)}_{\epsilon}(z)$ for $\epsilon<0$, b) $\varphi^{({\rm in},+)}_{\epsilon(z)}$ for $\epsilon>0$, 
c) $\varphi^{({\rm in},-)}_{\epsilon}(z)$ for $\epsilon>0$, d) $\varphi^{({\rm in},-)}_{\epsilon}(z)$ for $\epsilon<0$. In each case the incoming wave divides between two  outgoing waves.}
\label{FIG:instates}
\end{figure}

Similarly,  we can expand  the LLL field operator in terms of complete sets of parabolic cylinder functions. There are   two distinct ways of doing this.
Begin by defining
\bea
\varphi^{({\rm in},+)}_{\epsilon}(z)= \frac{1}{\pi^{1/4}} \Gamma(1/2-i\epsilon)U_{-i\epsilon}(z),\\
\varphi^{({\rm in},-)}_ {\epsilon}(z)= \frac{1}{\pi^{1/4}} \Gamma(1/2-i\epsilon)U_{-i\epsilon}(-z).
\eea
The label  ``in''  designates that  these wave-functions  describe states that have a simple description prior to their opportunity for tunnelling (see figure \ref{FIG:instates}) .
These ``in'' functions have been  normalized so that 
\be
\brak{\varphi^{({\rm in},\alpha)}_\epsilon}{\varphi^{({\rm in},\alpha')}_{\epsilon'}}= 2\pi \,\delta(\epsilon-\epsilon')\delta^{\alpha\alpha'}, \quad \alpha=\pm,
\ee
and they obey the LLL completeness relation
\be
\sum_{\alpha=\pm}  \int_{-\infty}^{\infty} \frac{d\epsilon}{2\pi}\varphi^{({\rm in},\alpha)}_\epsilon(z_1)\overline{\varphi^{({\rm in},\alpha)}_\epsilon(z_2)} = \frac 1{2\pi} \exp\left\{\frac 12 z_1\bz_2\right\}.
\ee
(Both  normalization and completeness  are easily established from the integral expression in the last line of  (\ref{EQ:cylinderdef}).)
Then we can set
\be
\widehat\psi(z) =
\int_{-\infty}^{\infty}\frac{d\epsilon}{2\pi} \left(({\widehat b}^{(\rm in)}_\epsilon)^\dagger \varphi^{({\rm in},+)}_{\epsilon}(z)+
 \widehat a^{({\rm in})} _\epsilon\varphi^{({\rm in},-)}_{\epsilon}(z)\right)
\ee
The ``in''  vacuum  is the  appropriate many-body state for our initial conditions. It is characterized physically by the condition that no particle is approaching the 2DEG from the empty single-particle states to the right, and that all the single-particle states  incoming  from the left are occupied. It is characterized mathematically by the  conditions
\be
\widehat a_\epsilon  \ket{0, {\rm in}}=0=\widehat b_\epsilon \ket{0, {\rm in}}, \quad \forall \epsilon.
\ee

\begin{figure}
\includegraphics[width=4.3in]{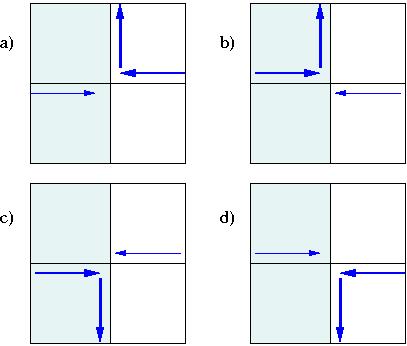}
\caption{\sl The ``out'' wavefunctions: a)  $\varphi^{({\rm out,ext})}_{\epsilon}(z)$ for $\epsilon>0$, b) $\varphi^{({\rm out,ext})}_{\epsilon}(z)$ for $\epsilon<0$. c) $\varphi^{({\rm out,int})}_{\epsilon}(z)$ for $\epsilon>0$,  d) $\varphi^{({\rm out,int})}_{\epsilon}(z)$ for $\epsilon<0$. In each case two  weaker incoming waves assemble  the outgoing wave.}
\label{FIG:outstates}
\end{figure}

The second set of functions  is
\bea
\varphi^{({\rm out,ext})}_{\epsilon}(z)= \frac{1}{\pi^{1/4}} \Gamma(1/2+i\epsilon)U_{i\epsilon}(iz),\\
\varphi^{({\rm out,int})}_ {\epsilon}(z)= \frac{1}{\pi^{1/4}} \Gamma(1/2+i\epsilon)U_{i\epsilon}(-iz).
\eea
They are also orthogonal
\be
\brak{\varphi^{({\rm out},\alpha)}_\epsilon}{\varphi^{({\rm out },\alpha')}_{\epsilon'}}= 2\pi \,\delta(\epsilon-\epsilon')\delta^{\alpha\alpha'}. \quad \alpha=\hbox{int, ext},
\ee
and  obey the LLL completeness relation
\be
\sum_{\alpha}  \int_{-\infty}^{\infty} \frac{d\epsilon}{2\pi}\varphi^{({\rm out},\alpha)}_\epsilon(z_1)\overline{\varphi^{({\rm out},\alpha)}_\epsilon(z_2)} = \frac 1{2\pi} \exp\left\{\frac 12 z_1\bz_2\right\}.
\ee
The labels ``ext'' and ``int'' indicate that the functions live mostly  in the exterior ($y>0$) and interior ($y<0$)  of the black hole. They decay rapidly in the other region (see figure \ref{FIG:outstates}).
In terms of these new functions we have
\be
\widehat\psi(z) =
\sum_{\alpha} \int_{-\infty}^{\infty}\frac{d\epsilon}{2\pi} \, {\widehat a}^{({\rm out},\alpha)}_\epsilon  \varphi^{({\rm out},\alpha)}_{\epsilon}(z)
\ee
The ``out''  operators  ${\widehat a}^{({\rm out},\alpha)}_\epsilon$ and $({\widehat a}^{({\rm out},\alpha)}_\epsilon)^\dagger$ create and annihilate particles that are simply described as excitations over the asymptotic na{\"i}ve vacuum in which every state in the region  $x<0$ is filled and every state in $x>0$ is empty.
For $\epsilon>0$ the operator $ {\widehat a}^{({\rm out,ext)}}_\epsilon$ annihilates a positive energy particle in the asymptotic region $y\gg 0$ outside the black hole. For  $\epsilon<0$ it annihilates a particle in the 2DEG and so creates a positive energy  hole in the same region.  For the $ {\widehat a}^{({\rm out,int)}}_\epsilon$ that act on states within the black hole  the roles of hole  creation and particle annihilation are reversed as the 2DEG consists of particles with positive energy.  To stress the causally disconnected character of the  interior and exterior regions, we will write  ``out'' vacuum as 
\be
\ket{0,{\rm out}}=\ket{0,{\rm out,ext}}\otimes \ket{0,{\rm out, int}}
\ee
with 
\bea
 {\widehat a}^{({\rm out,ext)}}_\epsilon \ket{0,{\rm out,ext}}&=&0, \qquad \epsilon>0\nonumber\\
({\widehat a}^{({\rm out,ext)}}_\epsilon)^\dagger \ket{0,{\rm out,ext}}&=&0, \qquad \epsilon<0,
\eea
and 
\bea
{\widehat a}^{({\rm out,int)}}_\epsilon \ket{0,{\rm out,int}}&=&0,\qquad \epsilon<0\nonumber\\
({\widehat a}^{({\rm out,int)}}_\epsilon)^\dagger \ket{0,{\rm out,int}}&=&0, \qquad \epsilon>0.
\eea

Comparing the two expressions for $\widehat \psi(z)$ gives us the Bogoliubov transformation
\bea
{\widehat a}^{({\rm in})}_\epsilon &=& \frac{\Gamma(\frac 12-i\epsilon)}{\sqrt{2\pi}}\left[ e^{-\epsilon \pi/2}e^{-i\pi/4} {\widehat a}_\epsilon^{(\rm out,int)}+
e^{\epsilon\pi/2} e^{i\pi/4} {\widehat a}^{(\rm out,ext)}_\epsilon\right],\\
{\widehat b}^{({\rm in})}_\epsilon &=& \frac{\Gamma(\frac 12+i\epsilon)}{\sqrt{2\pi}}\left[ e^{-\epsilon \pi/2}e^{i\pi/4} ({\widehat a}_\epsilon^{(\rm out,ext)})^\dagger+
e^{\epsilon\pi/2} e^{-i\pi/4} ({\widehat a}^{(\rm out,int)}_\epsilon)^\dagger\right].
\eea
Similarly 
\bea
{\widehat a}^{({\rm out,int})}_\epsilon &=& \frac{\Gamma(\frac 12+i\epsilon)}{\sqrt{2\pi}}\left[ e^{\epsilon \pi/2}e^{-i\pi/4} ({\widehat b}_\epsilon^{({\rm in})})^\dagger+
e^{-\epsilon\pi/2} e^{i\pi/4} {\widehat a}^{({\rm in})}_\epsilon\right],\\
{\widehat a}^{({\rm out,ext})}_\epsilon &=& \frac{\Gamma(\frac 12+i\epsilon)}{\sqrt{2\pi}}\left[ e^{\epsilon \pi/2}e^{-i\pi/4} {\widehat a}_\epsilon^{(\rm in )}+
e^{-\epsilon\pi/2} e^{i\pi/4} ({\widehat b}^{(\rm in )}_\epsilon)^\dagger\right].
\eea

From the Bolgoluibov transformation and the mathematical characterization of $\ket{0,{\rm in}}$  we find that
\be
\ket{0,{\rm in}}= N \exp \left\{i \int_0^\infty e^{-|\epsilon|\pi} \left[({\widehat a}^{({\rm out, ext})}_\epsilon)^\dagger {\widehat a}^{({\rm out, int})}_\epsilon
 +a^{({\rm out,ext})}_{-\epsilon} ({\widehat a}^{({\rm out,int})}_{-\epsilon})^\dagger\right] \frac{d\epsilon}{2\pi}\right\} \ket{0,{\rm out}},
 \ee
 where $N$ is a normalization factor. We have therefore exhibited the physical ground state as a sea of  particle-hole pairs correlated between the interior and exterior regions. 
 We now have the same formal situation as described in \cite{unruh-review}.
 If we trace out the ``unobservable'' interior of the black hole, we end up with  density matrix is of the form
 \be
\widehat  \rho = \sum_i e^{-2\pi |\epsilon(i)|} \ket{i,{\rm ext}}\otimes \bra{i,{\rm ext}},
 \ee
 where  $i$ labels the many-body state whose energy is $\epsilon(i)$. 
 However, unlike the situation in the Unruh-Rindler vacuum \cite{unruh76,unruh-review}  our system contains genuine radiation rather that a thermal bath. This is because the chiral character of the particles means that they can only flow outwards. 
 
 \section{Discussion}
 
 The effective space-time metric in which the chiral edge-mode fermions  move is 
 \be
 ds^2 =\frac{1}{v_{\rm edge}^2(y)} dy^2-dt^2
  \label{EQ:metric}
 \ee
 The quantization of  chiral fermions in such a background metric with general $v_{\rm edge}(y)$ has been carried out  in \cite{semenoff87}, although these authors did  not consider the effect of an event  horizon. 
 
 In our  case $v_{\rm edge}=\kappa  y$, $\kappa= \lambda/eB$, and a change to  an exterior  tortoise co-ordinate $y_*=\kappa^{-1} \ln (y) $  in (\ref{EQ:metric}) 
 leads to 
 \be
 ds^2= dy_*^2-dt^2.
 \ee
 The new coordinates  reveal that our  space-time is flat, but the singularity at the  horizon is not removed. It  has been pushed  to $y_*=-\infty$, and the interior of the black hole has become invisible.   A superfluid system with  this metric and   event horizon was  studied by Volovik in   \cite{volovik99}. He uses a WKB  analytic continuation method to compute the Bogoliubov coefficients, and  finds  the  same Hawking temperature as  our present calculation, but his non-chiral system has no actual radiation. The agreement in the temperature  is perhaps not surprising. It must be obvious from looking at the classical trajectories of our particles that there is some connection between our   2DEG problem and that of Landau-Zener tunneling through an avoided level crossing. Indeed, although the physics is superficially different, the Landau-Zener time-dependent Schr{\"o}dinger equation  is solved using the same families of parabolic cylinder functions that we have used \cite{zener}, and it is well known that an analytically continued form of the WKB approximation  obtains the correct   asymptotic Landau-Zener tunnelling probabilities \cite{landau}.   
 
 The most remarkable property of  the present model  is that the emitted radiation is exactly thermal.  There is no immediately obvious reason why the mathematical   properties of the parabolic cylinder functions  should  lead to this result. In a real  black hole  the emitted radiation is modified by grey-body factors in dimensions greater than two, but  that the hole can only be in equilibrium  with radiation at $T_{\rm Hawking}$ follows from the geometry of the Euclidean section of space-time  being asymptotically periodic  in imaginary time  \cite{gibbons-perry76}.    Does our  space-time geometry tacitly   force  a Euclidean temporal periodicity?
 
 We can write 
 \be
 ds^2 =\frac 1{\kappa^2 y^2} \left(dy^2-y^2 d(\kappa  t)^2\right)
 \label{EQ:rindler}
 \ee
 so, up to a conformal factor $\kappa^{-2} y^{-2}$, the metric is that of Rindler space whose  Euclidean section $t\mapsto i\tau$ has metric
 \be
 ds^2_{\rm Rindler} = y^2 d(\kappa \tau)^2 + dy^2.
 \label{EQ:euclidean_rindler}
 \ee
The absence of a conical singularity at $y=0$ in the manifold  described by (\ref{EQ:euclidean_rindler}) requires identifying $\kappa \tau \sim \kappa \tau+2\pi$ and so implies a  temperature $T=\hbar \kappa /2\pi$ --- which is exactly what the tunneling calculation gives.  However, given that  it blows up  at the point of interest, it  seems unreasonable to ignore the conformal factor, making this  argument  at most suggestive.

 \section{Acknowledgements}   This  project was supported by the National Science Foundation  under grant  DMR 09-03291.  I would like to thank Ted Jacobson for comments, and for drawing my attention to reference \cite{volovik99}.

 \end{document}